\documentclass[aps,prb,twocolumn,showpacs,superscriptaddress,floatfix]{revtex4}

\usepackage{amsmath}
\usepackage{epsfig,graphics}
\usepackage{braket}

\def\ba{\begin{array}}
\def\ea{\end{array}}

\begin{document}

\title{Intrinsic phonon decoherence and quantum gates in coupled lateral quantum dot charge qubits}

\author{Markus J. Storcz}
\email[]{markus.storcz@physik.lmu.de}
\author{Udo Hartmann}
\email[]{udo.hartmann@physik.lmu.de}
\affiliation{Physics Department, Arnold Sommerfeld Center for Theoretical Physics, and Center for NanoScience,\\
Ludwig-Maximilians-Universit\"at, Theresienstr. 37, D-80333 M\"unchen, Germany}
\author{Sigmund Kohler}
\affiliation{Institut f\"ur Physik, Universit\"at Augsburg, Universit\"atsstr. 1, 
D-86135 Augsburg, Germany}
\author{Frank K. Wilhelm}
\affiliation{Physics Department, Arnold Sommerfeld Center for Theoretical Physics, and Center for NanoScience,\\
Ludwig-Maximilians-Universit\"at, Theresienstr. 37, D-80333 M\"unchen, Germany}

\begin{abstract}

Recent experiments by Hayashi {\it et al.} [Phys. Rev. Lett. {\bf 91}, 226804 (2003)] demonstrate coherent
oscillations of a charge quantum bit (qubit) in laterally defined quantum dots. We study the
intrinsic electron-phonon decoherence and gate performance for the next step: a system of two coupled
charge qubits. The effective
decoherence model contains properties of local as well as collective decoherence.
Decoherence channels can be classified by their multipole moments, which leads to different low-energy spectra.
It is shown that due to the
super-Ohmic spectrum, the gate quality is limited by the single-qubit Hadamard gates. It can be
significantly improved, by using double-dots with weak tunnel coupling.
\end{abstract}


\maketitle

\section{Introduction}
In recent years, the experimental progress in analyzing transport properties
in double quantum dots \cite{vanderWiel2} has lead to the fabrication of
double dot structures with only one electron in the whole system \cite{Elzerman,Andi}. This
well-defined situation permits, although it is strictly speaking not necessary \cite{Hayashi},  to use quantum
dot systems as quantum bits (qubits).
In order to define qubits in lateral quantum dot (QD) structures, the two degrees of freedom,
spin and charge, are naturally used. For spin qubits \cite{Loss}, the information is encoded
in the spin of a single electron in one quantum dot, whereas for the charge 
qubit \cite{Blick1,Leonid1,vanderWiel1} the position of a single electron in a double dot 
system defines the logical states. Similar ideas can also be applied to charge states in 
Silicon donors \cite{Hollenberg}. Both realizations are interconnected: interaction and 
read-out\cite{Elzerman} of spin qubits are envisioned\cite{Loss} to be all-electrical 
and to make use of the charge degree of freedom. 

Although the promises of spin coherence in theory\cite{LatestLoss} and in bulk measurements 
\cite{Awschalom} are tremendous in the long run, it was
the good accessibility of the {\em charge} degrees of freedom which lead to a recent 
break-through \cite{Hayashi}, namely the demonstration of coherent oscillations in a quantum dot
charge qubit. In this experiment, three relevant
decoherence mechanisms for these charge qubits have been pointed out: a cotunneling 
contribution, the electron-phonon coupling, and $1/f$-noise or charge noise in the 
heterostructure defining the dots.

Recent theoretical results \cite{UdoFrank} predict that the cotunneling contribution
can be very small, provided that the coupling between the dots and the connected leads
is small. Thus, cotunneling is not a {\em fundamental} limitation. This, however, means that
initialization and measurement protocols different from those of Ref.~[\onlinecite{Hayashi}] are
favorable \cite{Elzerman}.

Other theoretical works \cite{Leonid2,Wu,Vorojtsov,Hu,thorwart_mucciolo} already describe the electron-phonon 
interaction for a single charge qubit in a GaAs/AlGaAs heterostructure. Moreover, also
electronic Nyquist noise in the gate voltages affects the qubit system \cite{Marquardt}. Note
that the physics of the electron-phonon coupling is different and less limiting in the unpolar material Si \cite{Gorman},
where the piezo-electric interaction is absent.

\section{Model}
In this article, we analyze the decoherence due to the electron-phonon coupling in GaAs, which is generally assumed
to be the dominant decoherence mechanism in a coupled quantum-dot setting.  The recent experimental analysis shows 
that the temperature dependence of the dephasing rate in the experiment
\cite{Hayashi} can be modeled with the Spin-Boson model and hence is compatible 
with this assumption \cite{Leggett}.
We develop a model along the lines of 
Brandes \textit{et al.} \cite{Brandes1,BrandesHabil} to describe the piezo-electric interaction
between electrons and phonons in lateral quantum dots. Thereby, we assume the distance between
the two dots to be sufficiently large and the tunnel coupling $\Delta$ to be relatively small, which
is a prerequisite for the validity of the model.
The Hamiltonian for a system of two double dots with a tunnel-coupling within the double dots and 
electrostatic coupling between them, see Fig.\ \ref{fig:coupled_dots}, can be expressed as \cite{BrandesHabil}
\begin{equation}
\hat{H}_{\rm total} = \hat{H}_{\rm sys} + \hat{H}_{\rm bath} + \hat{H}_{\rm int},
\end{equation}
where
\begin{eqnarray}
\label{eq:2qubits}
\hat{H}_{\rm sys} & = & - \sum_{i=1,2}\limits \frac{1}{2}\left( \varepsilon_i \hat{\sigma}_{z,i} + \Delta_i \hat{\sigma}_{x,i} \right)
- k \hat{\sigma}_{z,1}\otimes\hat{\sigma}_{z,2}\\
\hat{H}_{\rm bath} & = & \sum_q\limits \hbar \omega_q c^{\dagger}_q c_q
\end{eqnarray}
refer to the qubits and the heat bath, respectively. $q$ is the phonon wave number.
The system-bath interaction Hamiltonian $\hat{H}_{\rm int}$ depends on details of the setup such as
the crystalline structure of the host semiconductor
and the dot wave functions. 
We will distinguish between the two extreme cases of 
long correlation length phonons resulting in coupling of both qubits to a single phonon bath, 
 or two distinct phonon 
baths for short phonon correlation length. The former case is more likely \cite{Ulbrich} and applies to 
crystals which can be regarded as perfect and linear over the size of the sample, whereas the latter
case describes systems that are vstrained or disordered and double quantum dots in large geometrical separation. 
The correlation length has to be distinguished from the wave length: The former indicates, over which distances the phase of the phonon wave is maintained,{\it i.e.}, over which distance the description as  a genuine standing wave applies at all, whereas the latter indicates the internal length scale of the wave. 

\subsection{One common phonon bath}
\begin{figure}[t]
\centering
\includegraphics*[width=\columnwidth]{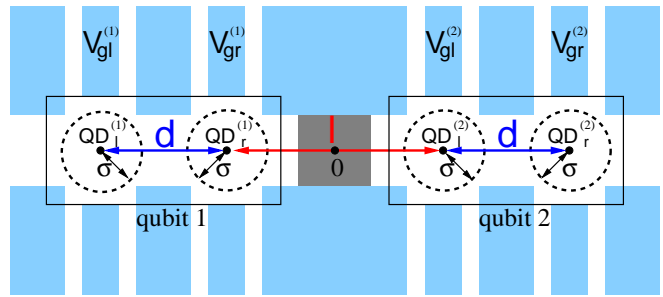}
\caption{(Colour online) Sketch of the two coupled identical charge qubits realized in a
lateral quantum dot structure. $d=100$~nm is the distance of the dot centers in one qubit, $l=200$~nm is
the distance between the right dot center of qubit 1 and the left dot center of qubit 2.
The width of the Gaussian wavefunction of an electron in each dot is $\sigma=5$~nm. The values chosen for the
distances $d$ and $l$ are slightly smaller than in experimental realizations \cite{Hayashi,Elzerman} in order to
provide a lower bound for the decoherence times. In principle, there could
be tunneling processes between both qubits, {\it i.e.}, the QDs two and three in the chain,
but we assume that the coupling is pinched off by applying appropriate gate voltages. The
gray box between the qubits indicates that there is no tunneling between the qubits.}
\label{fig:coupled_dots}
\end{figure}

In the case of a single phononic bath with a very long correlation length 
coupling to both charge qubits, $\hat H_{\rm int}$ can
be written as
\begin{eqnarray}
\hat{H}_{\rm int} & = & \sum_q\limits \frac{1}{2}\Big[ (\alpha_{q,1}+\beta_{q,1}+\alpha_{q,2}+\beta_{q,2})\hat{\mathbf{1}}_1\otimes\hat{\mathbf{1}}_2 + \nonumber \\
& & + (\alpha_{q,1}-\beta_{q,1})\hat{\sigma}_{z,1}\otimes\hat{\mathbf{1}}_2 + \nonumber \\
& & + (\alpha_{q,2}-\beta_{q,2})\hat{\mathbf{1}}_1\otimes\hat{\sigma}_{z,2}\Big] (c^{\dagger}_q + c_{-q})\ . \label{eq:initialelph}
\end{eqnarray}
The coefficients
$\alpha_{q,i}$ and $\beta_{q,i}$ describe the coupling of a localized
electron (one in each of the two double dot systems) to the phonon modes. They are given by
\begin{eqnarray}
\alpha_{q,i} & = & \lambda_q \langle l,i | e^{i\vec{q}\vec{x}} | l,i \rangle ,\\
\beta_{q,i} & = & \lambda_q \langle r,i | e^{i\vec{q}\vec{x}} | r,i \rangle ,
\end{eqnarray}
where the $|l,i\rangle$ and $|r,i\rangle$ denote the wavefunctions of the electrons in
the left or right dot of qubit $i$.\,We assume these wavefunctions to be two-dimensional
Gaussians centered at the center of the dot, as sketched in Figure~\ref{fig:coupled_dots}. These
states approximate the ground state in the case of a parabolic potential and small overlap
between the wavefunctions in adjacent dots. The
coefficient $\lambda_q$ is derived from the crystal properties \cite{BrandesHabil}.

Henceforth, we investigate the case of two identical qubits.  Due to the fact that
the relevant distances are arranged along the $x$-direction, we obtain the coupling coefficients
\begin{eqnarray}
\alpha_{q,1} & = & \lambda_q e^{iq\left(-{l}/{2}-d\right)} e^{-{q^2\sigma^2}/{4}},
\label{eq:c1}\\
\beta_{q,1} & = & \lambda_q e^{-{iql}/{2}} e^{-{q^2\sigma^2}/{4}}, \label{eq:c2}\\
\alpha_{q,2} & = & \lambda_q e^{{iql}/{2}} e^{-{q^2\sigma^2}/{4}}, \label{eq:c3}\\
\beta_{q,2} & = & \lambda_q e^{iq\left({l}/{2}+d\right)} e^{-{q^2\sigma^2}/{4}}\label{eq:c4}\ .
\end{eqnarray}
Here, $q$ is the absolute value of the wavevector $\vec{q}$. The second exponential function in each
line is the overlap between the two Gaussian wavefunctions.

This two-qubit bath coupling Hamiltonian is quite remarkable, as it does not fall into the two standard
categories usually treated in literature (see, {\it e.g.}, Refs.\ [\onlinecite{Kempe,Zanardi,pra_markus}] and references therein):
On the one hand, there is clearly only one bath and
each qubit couples to the bath modes with matrix elements of the same modulus, so
the noise between the qubits is fully correlated. On the other hand, the Hamiltonian does {\em not}
obey the familiar factorizing collective noise form $\hat{H}_\mathrm{SB,coll}=\hat{X}_{\rm
system}\otimes \hat{X}_{\rm bath}$. Such a form would lead
to a high degree of symmetry and thus protection from the noise coupling \cite{Zanardi,Kempe}, however, the Hamiltonian $H_{\rm int}$, eq. \ref{eq:initialelph}, cannot be factorized in such a bilinear form. It is hence
intriguing to explore where in between these cases the physics ends up to be. This is in particular
important for finally finding strategies to protect the qubits against decoherence, and for estimating the scaling
of decoherence in macroscopic quantum computers.

In order to obtain the dynamics of the reduced density matrix $\rho$ for the coupled qubits, {\it i.e.}, for
the degrees of freedom that remain after the environment is traced out, we apply Bloch-Redfield
theory \cite{Agyres,Weiss,Hartmann}.
It starts out from the Liouville-von Neumann equation $i\hbar\dot\rho =
[\hat H,\rho_\mathrm{tot}]$ for the total density operator.  A perturbational
treatment of the system-bath coupling Hamiltonian $\hat H_\mathrm{int}$
results in the master equation
\begin{equation}
\label{eq:br.basisfree}
\dot\rho = -\frac{i}{\hbar} [\hat H_\mathrm{sys},\rho]
- \frac{1}{\hbar^2}\int_0^\infty \!\!\! d\tau \ \mbox{tr}_B
[\hat H_\mathrm{int}, [\tilde H_\mathrm{int}(-\tau),\rho\otimes\rho_B ]] ,
\end{equation}
where $\rho_B = \exp(-\beta \hat H_B)/Z$ denotes the equilibrium density matrix of
the bath.  Evaluating the trace over all bath variables, $\mbox{tr}_B$, and
decomposing the reduced density operator into the eigenbasis of the
unperturbed system Hamiltonian, we obtain \cite{Blum,Weiss}
\begin {equation}
\dot{\rho}_{nm} = - i\omega_{nm}\rho_{nm} - \sum_{k,\ell} R_{nmk\ell}\rho_{k\ell} , \label{eq:br}
\end {equation}
where $\omega_{nm} = E_n-E_m$.
The first term on the right hand side describes the unitary evolution and the
Redfield relaxation tensor $R_{nmk\ell}$ incorporates the decoherence effects. It is given by
\begin {equation}
R_{nmk\ell} = \delta_{\ell m}\sum_r\Gamma^{(+)}_{nrrk} + \delta_{nk}\sum_r\Gamma^{(-)}_{\ell rrm} - \Gamma^{(-)}_{\ell mnk} - \Gamma^{(+)}_{\ell mnk}\textrm{,}
\end {equation}
where the rates $\Gamma^{(\pm)}$ are determined by Golden Rule expressions\cite{Blum,Weiss},
see Eqs.~\eqref{Gamma+} and \eqref{Gamma-}, below.
The Redfield tensor and the time evolution of the reduced density matrix are evaluated numerically
to determine the decoherence properties of the system due to a weak electron-phonon coupling.
Note that in addition, Ohmic electronic noise can be taken into account by employing the
spectral function\cite{governale} $J_\Sigma (\omega) = J_{\rm Ohmic}(\omega) + J (\omega)$, 
where $J(\omega)$ contains only the phonon contribution. It
is also possible to take $1/f$-noise in the quantum dot system into account in the same way.
The $1/f$-noise essentially determines the magnitude of the dephasing part of the
decoherence. Thus, it is in turn possible to impose for the zero frequency component $J(0)$ the
experimental value of the dephasing rates or a value from a microscopic model \cite{Falci}. 
However, in many cases it turns out to be non-Markovian and/or non-Gaussian, leading to non-exponential decay, 
which can neither be described by Bloch-Redfield theory nor parameterized by a single rate.

In order to compute the rates, the electron-phonon interaction Hamiltonian has first to be taken from the localized
representation to the computational basis, which is straightforward. To compute
Bloch-Redfield rates, it is necessary to rotate into the eigenbasis of the system.
After  this basis change, the spectral densities $J_{\ell mnk}(\omega)$ are calculated
along the lines of Ref.~[\onlinecite{BrandesHabil}] as
\begin{equation} \label{eq:specf}
J_{\ell mnk}(\omega ) = \langle \left( B^{-1} C B \right)_{\ell m}\left( B^{-1} C B \right)_{nk}\rangle_{q} ,
\end{equation}
where $B$ is the matrix for the basis transformation from the computational basis \{$|00\rangle$, $|01\rangle$,
$|10\rangle$, $|11\rangle \}$ to the eigenbasis of the system and $\langle \cdot \rangle_q$ denotes an averaging
over all phonon modes $q$ with frequency $\omega$. The matrix $C$ is diagonal in the
computational basis,  $C={\rm diag }(\alpha_{q,1}-\beta_{q,1}+\alpha_{q,2}-\beta_{q,2}, \alpha_{q,1}-\beta_{q,1}+\alpha_{q,2}-\beta_{q,2}, \alpha_{q,1}-\beta_{q,1}+\alpha_{q,2}-\beta_{q,2}, \alpha_{q,1}-\beta_{q,1}+\alpha_{q,2}-\beta_{q,2})$.

\begin{figure}[t]
\centering
\includegraphics*[width=0.5\columnwidth]{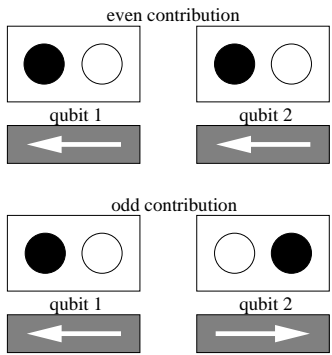}
\caption{Illustration of the even (top) and odd (bottom) contributions to the total rates. 
Filled circles indicate occupied dots. For long-wavelength modes, the energy shifts induced by underlying 
phonons in the two dots add up coherently in the even case but cancel in the odd case. Note, 
that moving charges from the black to the white dots changes the dipole moment in the even but not
in the odd case.}
\label{fig:diquad}
\end{figure}

\begin{figure}[t]
\centering
\includegraphics*[width=\columnwidth]{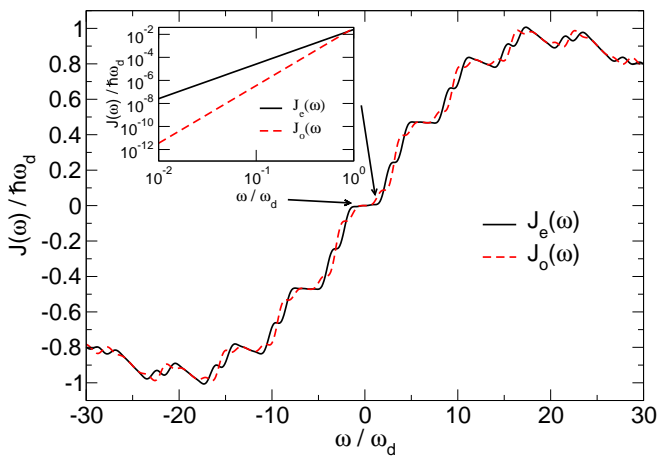}
\caption{(Colour online) Spectral functions $J_{e,o}(\omega)$ in the case of one common phonon bath 
for the fixed parameters $c_s = 5000$~m/s, 
$g = 0.05$, $d = 100$~nm, $l = 200$~nm and $\sigma = 5$~nm. Inset: zoom
for small frequencies.}
\label{fig:J1}
\end{figure}

The explicit derivation shows that it is most convenient to split the total spectral function
$J_{\ell mnk}(\omega)$ [see Eq.~(\ref{eq:specf})] into odd and even components
\begin{equation}
J_{\ell mnk}(\omega ) = e_{\ell mnk} J_e(\omega ) + o_{\ell mnk} J_o(\omega ) \ ,
\end{equation}
where the prefactors $e_{\ell mnk}$ and $o_{\ell mnk}$ of the even/odd part of the spectral function
are matrix elements coming from the basis change from the computational
basis to the eigenbasis of the system and
\begin{equation}
J_{e/o}(\omega)=\frac{\pi}{4}\sum_q |\alpha_{q,1}-\beta_{q,1}\pm\alpha_{q,2}\mp\beta_{q,2} |^2\delta(\omega-\omega_q) .
\end{equation}
They evaluate to
\begin{widetext}
\begin{equation}
J_{e,o}(\omega ) = \frac{\pi \hbar \omega g}{4} \Bigg[ 2-
2\frac{\omega_d}{\omega}\sin\left(\frac{\omega}{\omega_d}\right)\mp\frac{\omega_l}{\omega}\sin\left(\frac{\omega}{\omega_l}\right)\pm2\frac{\omega_{l+d}}{\omega}\sin\left(\frac{\omega}{\omega_{l+d}}\right)\mp\frac{\omega_{l+2d}}{\omega}\sin\left(\frac{\omega}{\omega_{l+2d}}\right)\Bigg]
e^{-{\omega^2}/{2\omega_c^2}} ,
\end{equation}
\end{widetext}
where $g=0.05$ is the dimensionless electron-phonon coupling strength
for the commonly used material GaAs \cite{Brandes1,BrandesHabil} and $c_S$ the speed of sound. 
The different frequencies represent the distances in the system: $\omega_d = c_s/d$,
$\omega_l = c_s/l$, $\omega_{d+l} = c_s/(d+l)$ and $\omega_{2d+l} = c_s/(2d+l)$, and $\omega_c=c_s/\sigma$. 
This structure can be
understood as follows: The electron-phonon interaction averages out if the phonons are rapidly oscillating within a
dot, {\it i.e.} if the wavelength is much shorter than the dot size --- this provides the high-frequency cutoff
at $\omega_c$. On the other hand, long-wavelength 
phonons do not contribute to decoherence between dots $i$ and $j$, if the wavelength is much longer than their
separation because then, the energy shift induced by the phonon displacement will only lead to a global phase. Furthermore, we
can approximate the leading order at low frequencies as 
\begin{eqnarray}
J_e(\omega) & = & \frac{2 \pi\hbar g d^2}{3c_s^2}   \omega^3 +\mathcal{O}(\omega^5)\label{eq:lowfreq1} , \\
J_o(\omega) & = & \frac{\pi\hbar g (l^2d^2+2ld^3+d^4)}{10c_s^4}\omega^5+\mathcal{O}(\omega^7)  \label{eq:lowfreq2} .
\end{eqnarray}
This different power-laws $\omega^3$ to $\omega^5$ can be understood
physically as illustrated in Figure \ref{fig:diquad}. 
``Even'' terms are the natural extension of the one-qubit electron-phonon coupling, adding up coherently
between the two dots. In the ``odd'' channel, the energy offset induced in one qubit is, for long wavelengths, 
cancelled by the offset induced in the other qubit. Thus, shorter wavelenghts are required for finding a 
remaining net effect. An alternative point of view is the following: The distribution of the two charges can be parameterized by a
dipole and a quadrupole moment. The ``even'' channel couples to the dipole
moment of the charge configuration similar to the one-qubit case. The ``odd'' channel couples to the
quadrupole moment alone (see Figure~\ref{fig:diquad}). Thus, it requires shorter wavelengths and consequently is 
strongly supressed at low frequencies. This explains the different low-frequency behavior 
illustrated for realistic parameters in Figure~\ref{fig:J1}.
Thus, we can conclude that for small frequencies the odd processes are suppressed by symmetry --- even 
beyond the single-dot supression and the suppression of asymmetric processes.

With these expressions for the spectral densities, one can proceed as in
Ref. [\onlinecite{pra_markus}] and determine the rates that constitute
the Redfield tensor to read
\begin{eqnarray}
\Gamma^{(+)}_{\ell mnk} & = & \frac{J_{\ell mnk}(\omega_{nk})}{2\hbar}
\left[\coth\left(\frac{\hbar\omega_{nk}}{2 k_B T}\right) - 1\right] , \label{Gamma+} \\
\Gamma^{(-)}_{\ell mnk} & = & \frac{J_{\ell mnk}(\omega_{\ell m})}{2\hbar}
\left[\coth\left(\frac{\hbar\omega_{\ell m}}{2 k_B T}\right) + 1\right] . \label{Gamma-}
\end{eqnarray}
For $\omega_{ij} \rightarrow 0$, these rates vanish due to the super-Ohmic form of the
bath spectral function.  From this, we find the time evolution of the coupled qubit
system and finally also the gate quality factors.

\subsection{Two distinct phonon baths}

When each qubit is coupled to its own phononic bath, the part of the Hamiltonian that describes
the interaction with the environment $H_{\rm int}$ is given by
\begin{eqnarray}
\hat{H}_{\rm int} & = & \sum_{q_{1}}\limits \frac{1}{2}\Big[ (\alpha_{q_{1}}+\beta_{q_{1}})\hat{\mathbf{1}}_1 + (\alpha_{q_{1}}-\beta_{q_{1}})\hat{\sigma}_{z,1} \Big] \nonumber \\
& & \times (c^{\dagger}_{q_{1}} + c_{-q_{1}}) \otimes \hat{\mathbf{1}}_2 + \nonumber \\
& & + \sum_{q_{2}}\limits \frac{1}{2}\Big[ (\alpha_{q_{2}}+\beta_{q_{2}})\hat{\mathbf{1}}_2 + (\alpha_{q_{2}}-\beta_{q_{2}})\hat{\sigma}_{z,2} \Big] \nonumber \\
& & \times (c^{\dagger}_{q_{2}} + c_{-q_{2}}) \otimes \hat{\mathbf{1}}_1 \ .
\end{eqnarray}
This scenario can be realized in different ways: One can split the crystal into two pieces by an
etched trench. Alternatively, if there is lattice disorder and/or strong nonlinear effects, the phonons
between the dots may become uncorrelated.

The calculation of the coupling coefficients works in a similar way, but there are two
different indices $q_1$ and $q_2$ to represent the phononic baths of each qubit
\begin{eqnarray}
\alpha_{q_{1}} & = & \lambda_{q_{1}} e^{iq_1\left(-{l}/{2}-d\right)} e^{-{q_1^2\sigma^2}/{4}}, \\
\beta_{q_{1}} & = & \lambda_{q_{1}} e^{-{iq_1l}/{2}} e^{-{q_1^2\sigma^2}/{4}}, \\
\alpha_{q_{2}} & = & \lambda_{q_{2}} e^{{iq_2l}/{2}} e^{-{q_2^2\sigma^2}/{4}}, \\
\beta_{q_{2}} & = & \lambda_{q_{2}} e^{iq_2\left({l}/{2}+d\right)} e^{-{q_2^2\sigma^2}/{4}}\ .
\end{eqnarray}

The expression
for the spectral functions $J_{\ell mnk}(\omega)$ turns out to be exactly the same as the one in the last section
with the only difference that instead of $\alpha_{q,i}$, the coupling between
electrons and phonons is now expressed as $\alpha_{q_{i}}$ (with $i=1,2$ for both qubits).
Therefore, in order to obtain the spectral density $J_{\ell mnk}(\omega)$,
one has to average over two distinct baths, {\it i.e.}
\begin{equation}
J_{\ell mnk}(\omega ) = \langle \left( B^{-1} C B \right)_{\ell m}\left( B^{-1} C B \right)_{nk}\rangle_{q_{1},q_{2}} .
\end{equation}
Again, we find two different functions that we name in the same way as in the
previous section, $J_e(\omega)$ and $J_o(\omega)$, which are given by
\begin{widetext}
\begin{equation}
J_{e,o}(\omega ) = \frac{\pi \hbar \omega g}{4} \Bigg[ 2-2\frac{\omega_d}{\omega}\sin\left(\frac{\omega}{\omega_d}\right) \mp 2\left(
\frac{\omega_{\frac{l}{2}}}{\omega}\sin\left(\frac{\omega}{\omega_{\frac{l}{2}}}\right)-\frac{\omega_{d+\frac{l}{2}}}{\omega}\sin\left(\frac{\omega}{\omega_{d+\frac{l}{2}}}\right)\right)^2 \Bigg]
e^{-{\omega^2}/{2\omega_c^2}} \ .
\end{equation}
\end{widetext}
The prefactors from the basis change also enter the expressions for the rates in the
same way as in the last section.
\begin{figure}[t]
\centering
\includegraphics*[width=\columnwidth]{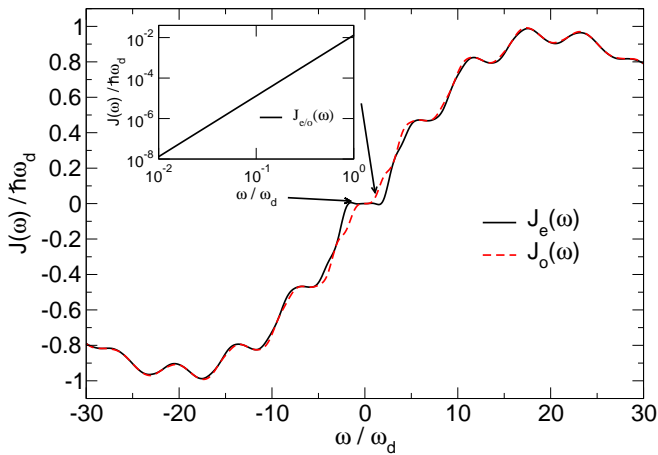}
\caption{(Colour online) Spectral functions $J_{e,o}(\omega)$ in the case of two distinct phonon baths 
for the fixed parameters $c_s = 5000$~m/s, 
$g = 0.05$, $d = 100$~nm, $l = 200$~nm and $\sigma = 5$~nm. Inset: magnification for small
frequencies.}
\label{fig:J2}
\end{figure}
The spectral functions $J_{e,o}(\omega)$ are plotted in Figure~\ref{fig:J2}; the inset depicts
the proportionality to $\omega^3$ for small frequencies.

\section{Golden Rule Rates}

We proceed as in Ref.~[\onlinecite{pra_markus}] and determine
the Golden rule rates that govern the Redfield tensor.  Thereby, we find both
the time evolution of the coupled system and the gate quality factors.

Let us first discuss the impact of this particular bath coupling on the dephasing and relaxation
rates. The decoherence rates, {\it i.e.}, the relaxation and dephasing rates, are defined according
to $\Gamma_R = - \sum_n \Lambda_n$, where $\Lambda_n$ are the eigenvalues of the matrix composed of
the elements $R_{n,n,m,m}$, $n,m=1,\ldots,4$, and $\Gamma_{\varphi_{nm}}=-\mbox{Re}R_{n,m,n,m}$ for non-degenerate
levels $\vert \omega_{nm} \vert > \vert R_{n,m,n,m} \vert$ and in the absence of
Liouvillian degeneracy, $\vert \omega_{nm} -\omega_{kl}\vert > \vert R_{a,b,c,d} \vert\quad a,b,c,d,\in\lbrace k,l,m,n\rbrace$, respectively \cite{governale}.

As a reference point, we study the rates in the uncoupled case. In this case, and in the absence of
degeneracies between the qubits, there is a clear selection rule that the environment only 
leads to single-qubit processes, {\it i.e.}, decoherence can be treated at completely separate 
footing. As a result, all rates are identical between the qubits. To make this obvious, we 
rewrite the original Hamiltonian in the one-bath case, combining Eq.~(\ref{eq:initialelph}) 
with eqs.~(\ref{eq:c1})--(\ref{eq:c4}) as
\begin{eqnarray}
\hat{H}_{\rm int}& = &\sum_q\limits \Bigg[-2i e^{-q^2\sigma^2/4}\sin\left(\frac{qd}{2}\right) \Big(e^{-iq(l+d)/2}\hat{\sigma}_{z,1} +\nonumber \\
& & +e^{iq(l+d)/2}\hat{\sigma}_{z,2}\Big)+ E_0\hat{\mathbf 1}\Bigg] \left(c^\dagger_{q} + c_{-q} \right)
\end{eqnarray}
which --- besides a phase factor which is meaningless for single-qubit transitions --- is identical to
the standard electron-phonon Hamiltonian for double dots \cite{BrandesHabil}. 

\begin{figure*}[t]
\centering
\includegraphics*[width=1.7\columnwidth]{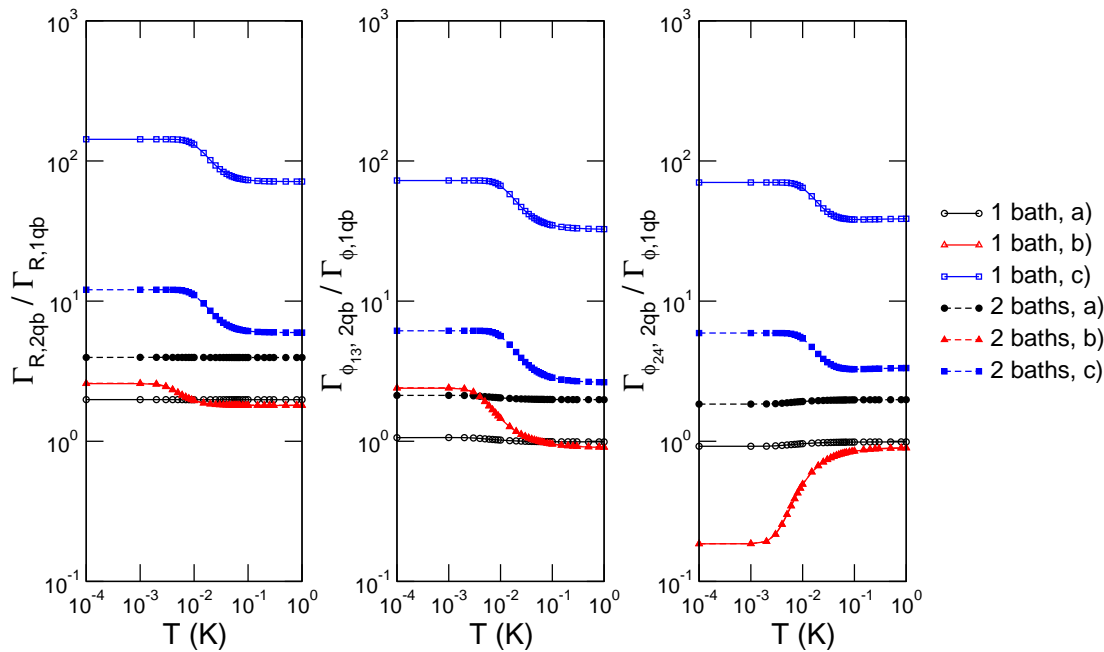}
\caption{(Colour online) Temperature dependence of the relaxation and dephasing rates normalized by the single-qubit
relaxation and dephasing rates. The two-qubit relaxation rate is given by the trace of the
relaxation part of the Redfield tensor in secular approximation. The energy scales for the two-qubit
transitions $1 \leftrightarrow 3$ and $2 \leftrightarrow 4$ are comparable to the single qubit
energy scale, the characteristic qubit energies are $E_s=(1/8)$~GHz.
The different cases are (a) $\varepsilon_1=\Delta_1 = (1/40) E_s$, $\varepsilon_2=\Delta_2 = -(21/40) E_s$,
and coupling energy $K = 0$, (b) $\varepsilon_1=\Delta_1 = -(1/2) E_s$, $\varepsilon_2=\Delta_2 = -(21/40) E_s$,
and $K = 0$), and (c) $\varepsilon_1=\Delta_1 = -(1/2) E_s$, $\varepsilon_2=\Delta_2 = -(1/2) E_s$
and $K = 10 E_s$. Note that cases (a) and (b) model uncoupled qubits, especially for case (a) the
overall relaxation rate for the two-qubit system is approximately twice the single-qubit relaxation
rate when calculated for the dominating larger energy scale of the two-qubit system ($\varepsilon_2=\Delta_2 = -(21/40) E_s$).
}
\label{fig:rates}
\end{figure*}
Figure~\ref{fig:rates} shows the temperature dependence of the energy relaxation rate $\Gamma_{R}$ and
the two dephasing rates $\Gamma_{\phi_{13}}$ and $\Gamma_{\phi_{24}}$ compared to the
single qubit relaxation and dephasing rates. In this notation, $\Gamma_{\phi_{ij}}$ is the rate at
which a superposition of energy eigenstates $i$ and $j$ is decays into a classical mixture. We considered the following
three cases, characterized by values on the matrix element relative to a characteristic system energy 
scale $E_s$: (a) large difference of  the $\varepsilon_i$ and $\Delta_i$ ($i = 1,2$) between both qubits and
no coupling between the qubits ($\varepsilon_1=\Delta_1 = (1/40) E_s$, $\varepsilon_2=\Delta_2 = -(21/40) E_s$
and coupling energy $K = 0$), (b) small asymmetry between the parameters for both qubits
and no coupling ($\varepsilon_1=\Delta_1 = -(1/2) E_s$, $\varepsilon_2=\Delta_2 = -(21/40) E_s$
and $K = 0$), and (c) without asymmetry between the qubits and a rather strong coupling between
the qubits ($\varepsilon_1=\Delta_1 = -(1/2) E_s$, $\varepsilon_2=\Delta_2 = -(1/2) E_s$
and $K = 10 E_s$). One generally would expect a different value of the distance between the dot centers in the qubits $d$,
when the tunneling coupling is varied. However, in our case of the dot
wavefunctions which overlap only in their Gaussian tails, this
effect is very small (below $1$~nm for a change in the tunneling amplitude $\Delta$ of approximately $\sim (1/2) E_s$)
for the lengthscales that we are considering. Note, that in Ref. \onlinecite{Andi} a substantial change of $\Delta$ over more than an order of magnitude was obtained experimentally by a rather mild adjustment of the gate
voltage, so it is consistent that a small change of $\Delta$ can be achieved by a tiny adjustment.
Therefore the value $d=100$~nm is used 
for the electron-phonon coupling encoded in $J_e$ and $J_o$ in all cases.

For case (a), we find that all rates are for all temperatures larger than the single qubit rates, as one would expect
\cite{Duer}. In more detail,
for the single bath case, the ratio of the relaxation rates is approximately $1.9$, the ratio of the single-qubit
dephasing rate and the two-qubit dephasing rate $\Gamma_{\phi_{24}}$ is around $0.9$ and for the 
dephasing rate $\Gamma_{\phi_{13}}$, the ratio
is $1.0$. The behaviour of the even and odd parts of the spectral function in the single bath case can
be explained from the spectral function Fig.~\ref{fig:J1}, for small $\omega$ one finds that $J_o<J_e$.
For the case of large frequencies, however, the even part of the spectral functionincreases and even dominate beyond the threshold $\omega \gtrsim \omega_d$.
Overall, it is found that in the case of a single-bath the decoherence effects are significantly suppressed
compared to the two-bath scenario.
For the two-bath case, the ratios are for the relaxation rates approximately $3.9$, for the
dephasing rate $\Gamma_{\phi_{24}}$ around $1.9$ and for the dephasing rate $\Gamma_{\phi_{13}}$ it 
is $2.0$. Note that for the two-bath case $J_e<J_o$ always and for the case where both tunnel
matrix elements in the Hamiltonian vanish, the rate vanishes, too.

After decreasing the asymmetry between the two qubits as in case (b), the rates decreased but are still
comparable with the single qubit rates, besides the last dephasing rate $\Gamma_{\phi_{24}}$. This can be 
understood, if one considers the energy spectrum of the eigenvalues of the system Hamiltonian. 
In cases (a) and (b) there is significant difference between the qubits, so it is straightforward to 
map the two-qubit rates onto the corresponding single qubit rates and they are largely determined by
single-qubit physics. 
In case (c), we consider a fully symmetric case in the qubit parameters, but with a finite and large coupling
between the qubits. This coupling lifts the degeneracy but makes the rate a generic two-qubit rate which
belongs to a relatively robust transition with small transition matrix elements for the
single bath case. At high temperatures,
these symmetry-related effects wash out as discussed in Ref.~[\onlinecite{Praktikanten}]. However,
the high-temperature rates do {\em not} coincide with the single-qubit rates, as the underlying energy scales are still different and in generally larger for the two-qubit situation.

Overall, the ratio of the two-qubit and single-qubit relaxation
rates decreases for increasing temperature due to the reduction of correlation effects in the
double dot system, besides case c), where a symmetry based on the underlying Hamiltonian 
becomes important.

\section{Quantum gate performance}

For the characterization of the quantum gate performance of this two-qubit system, it is necesssary to
introduce suitable quantifiers. Commonly, one employs the four gate quality factors introduced
in Ref.~[\onlinecite{pcz}]; fidelity $\mathcal{F}$, purity $\mathcal{P}$, quantum
degree $\mathcal{Q}$, and entanglement capability $\mathcal{C}$ to chararcterize
a gate operation within a hostile enviroment.

The fidelity, {\it i.e.}, the overlap between the ideal propagator and the simulated time evolution
including the decoherence effects, is defined as
\begin {equation}
\label{fidelity}
\mathcal{F} = \overline {\bra{\Psi_{\rm in}}\hat{U}^\dagger \rho_{\rm out}\hat{U} \ket{\Psi_{\rm in}}}\textrm{,}
\end {equation}
where the bar indicates an average over a set of 36 unentangled input states $\ket{\Psi_{\rm in}} =
\ket{\psi_i}\ket{\psi_j}$, with $i,j=1,\ldots,6$.  The 6 single-qubit states
$\ket{\psi_i}$ are chosen such that they are symmetrically distributed over the Bloch sphere, 
\begin{equation}
\ket{\psi_1}=\ket{0},\quad
\ket{\psi_2}=\ket{1},\quad
\ket{\psi_{3,\ldots,6}} =
\frac{\ket{0}+e^{i\phi}\ket{1}}{\sqrt{2}}
\end{equation}
where $\phi=0,\pi/2,\pi,3\pi/2$.
Here,  $\hat U$ is the ideal unitary time evolution for the given gate, and $\hat{\rho}_{\rm out}$ is the reduced density matrix
resulting from the simulated time evolution. A perfect gate reaches a fidelity of unity.
The purity $\mathcal{P}$ measures the strength of the decoherence effects,
\begin {equation}
\mathcal{P} = \overline {\mbox{tr}( \rho_{\rm out}^2)}\textrm{.}
\end {equation}
Again, the bar indicates the ensemble average. A pure state returns unity and for a mixed state the purity can drop to
a minimum given by the inverse of the dimension of the system Hilbert space, \textit{i.e.} 1/4 in our case.
\begin{figure*}[t]
\centering
\includegraphics*[width=1.8\columnwidth]{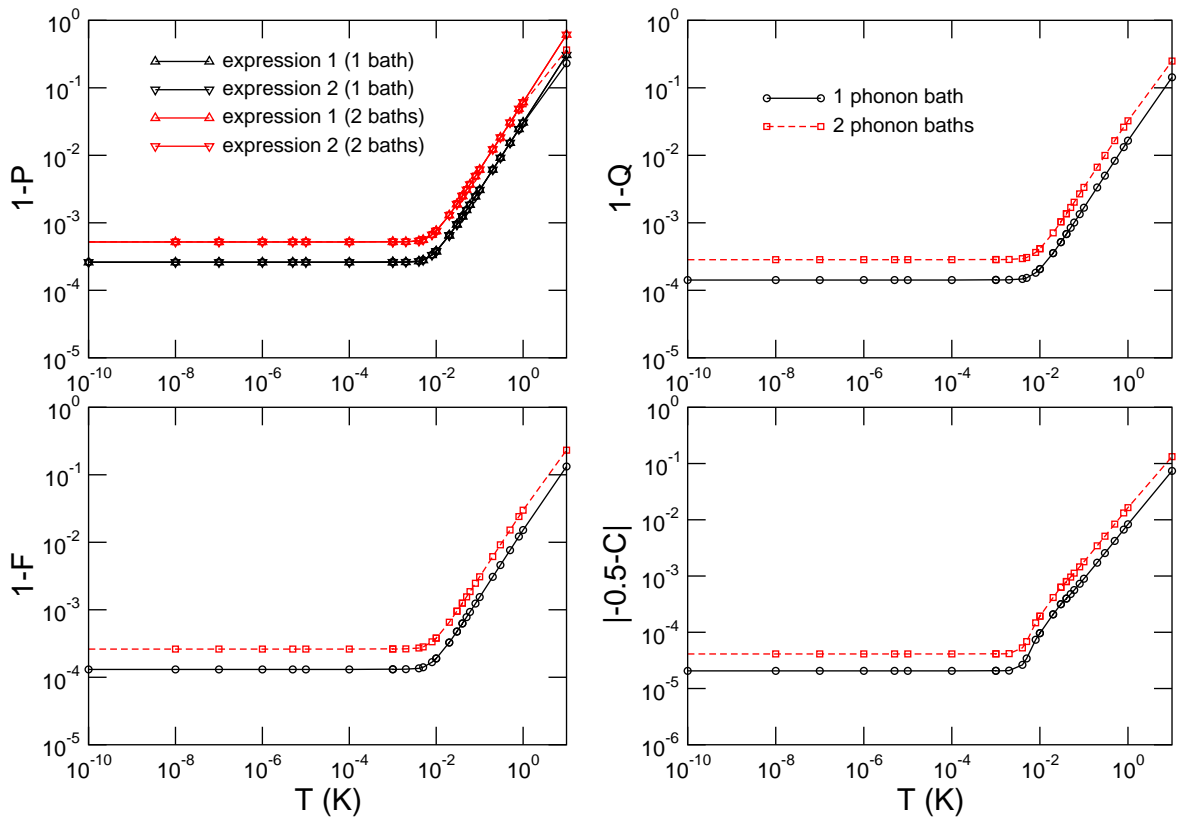}
\caption{(Colour online) Temperature dependence of the deviation of the four gate quality factors from their ideal values for the {\sc cnot} gate.
The decoherence due to phonons is taken into account. The black line shows the results for a single phonon bath and
the red line is for two phononic baths. The characteristic qubit energies are $E_s=1/4$~GHz
and the tunnel amplitudes are $\Delta_i = E_s$ ($i = 1,2$) due to the spacing of the double dots. 
In the curves for the deviation of the purity, we included lines for the analytical expressions 1 from
Eq.~(\ref{eq:exp1}) and 2 from Eq.~(\ref{eq:exp2}).}
\label{fig:phonons}
\end{figure*}

\begin{figure*}[t]
\centering
\includegraphics*[width=1.8\columnwidth]{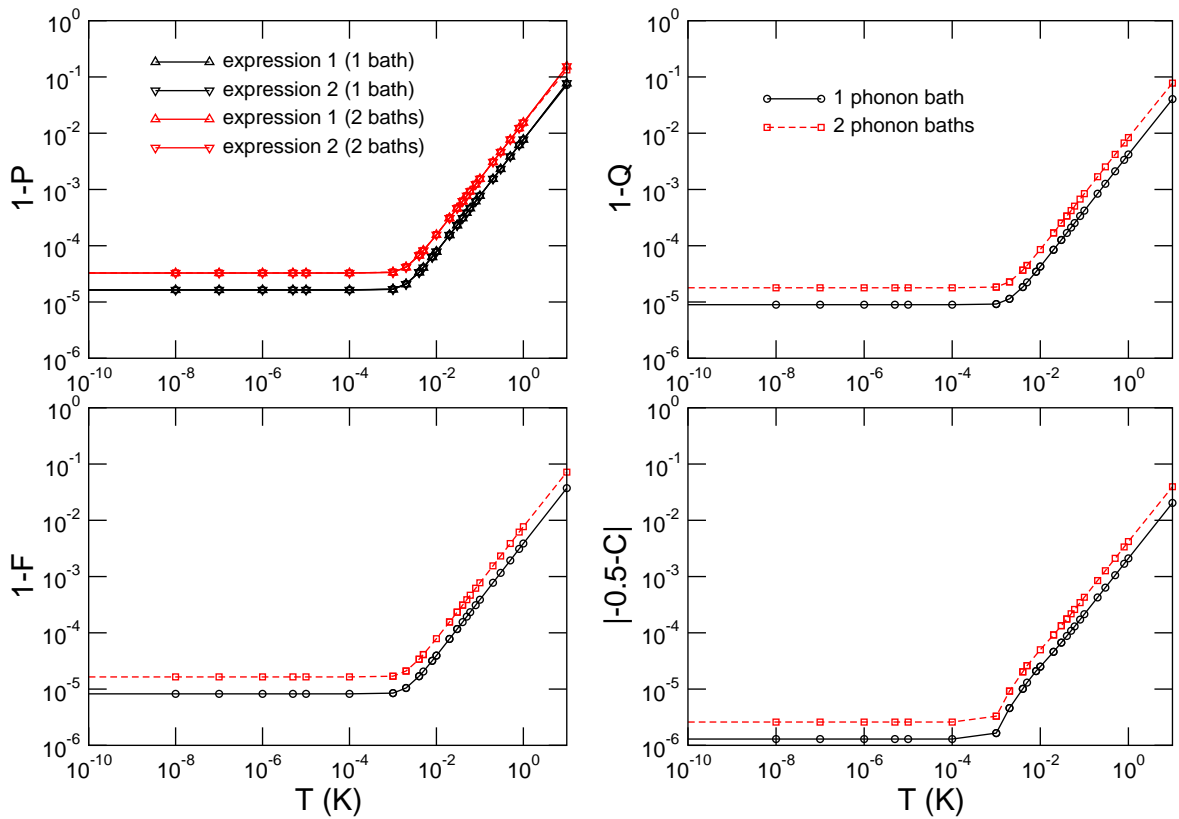}
\caption{(Colour online) Temperature dependence of the deviation of the four gate quality factors from their ideal values for the {\sc cnot} gate.
The decoherence due to phonons is taken into account. The black line shows the results for a single phonon bath and
the red line is for two phonon baths. The characteristic qubit energies are $E_s=1/4$~GHz and the tunnel amplitude during the Hadamard operation
on the second qubit is $\Delta_2 = 1/4 E_s$, {\it i.e.}, a factor $4$ smaller than in Figure~\ref{fig:phonons}. In the curves 
for the deviation of the purity, we included lines for the analytical expressions 1 from
Eq.~(\ref{eq:exp1}) and 2 from Eq.~(\ref{eq:exp2}).}
\label{fig:phonons_sc}
\end{figure*}

If the density operator $\rho$ describes an almost pure state, \textit{i.e.},
if the purity is always close to the ideal value 1, it is possible to estimate
the purity loss during the gate operation from its decay rate along the lines
of Ref.~[\onlinecite{Fonseca04}]. Thereby, one
first evaluates the decay of $(d/dt)\mbox{tr}\rho^2$ for an arbitrary pure
qubit state $\rho = |\psi\rangle\langle\psi|$.  From the basis-free version of the master
equation~\eqref{eq:br.basisfree}, follows straightforwardly
\begin{equation}
\label{rho2dot}
\frac{d}{dt}\mbox{tr} \rho^2 =
- \frac{2}{\hbar^2}\int_0^\infty d\tau \ \mbox{tr}_{S+B}
[\hat H_\mathrm{int}, [\tilde H_\mathrm{int}(-\tau),\rho\otimes\rho_B ]]\rho\textrm{.}
\end{equation}
By tracing out the bath variables, we obtain an expression that contains only
qubit operators and bath correlation functions.  This depends  on the state $|\psi\rangle$ via the
density operator.  Performing the
ensemble average over all pure states as described in the Appendix
\ref{app:average}, we obtain
\begin{equation} \label{eq:exp1}
\dot{\mathcal{P}}
= \frac{2}{\hbar^2(N+1)} \int_0^\infty d\tau \ \mbox{tr} \langle [\hat H_\mathrm{int} ,
\tilde H_\mathrm{int}(-\tau)]_+\rangle_{B,\mathrm{eq}} , 
\end{equation}
where $N=4$ denotes the dimension of the system Hilbert space of the two
qubits.  We have used the fact that $\mbox{tr} \hat H_\mathrm{int}=0$.
Although the discrete and set of states employed in the numerical computation is obviously different
from the set of all pure states, we find that both ensembles provide essentially the same results for
the purity.

If the bath couples to a good quantum number, {\it i.e.}, for
$[\hat H_\mathrm{sys}, \hat H_\mathrm{int}] = 0$, the system operator contained in
the interaction picture operator $\tilde H_\mathrm{int}(-\tau)$ remains
time-independent.  Then, the $\tau$-integration in \eqref{eq:exp1}
is effectively the Fourier transformation of the symmetrically ordered bath
correlation function in the limit of zero frequency.  Thus, we obtain
\begin{equation}
\label{Pdot0}
\dot{\mathcal{P}} = -\frac{2}{N+1} \lim_{\omega\to 0} \sum_i J_i(\omega)
\coth\frac{\hbar\omega}{2kT} ,
\end{equation}
where
\begin{equation}
J_i(\omega) = \frac{\pi}{4}\sum_q |\alpha_{q,i}-\beta_{q,i}|^2
\delta(\omega-\omega_q)
\end{equation}
denotes the spectral density of the coupling between qubit $i$ and the heat
bath(s).

In the present case of a super-Ohmic bath, the limit $\omega\to 0$ results for
the coupling to a good quantum number in $\dot{\mathcal{P}}=0$.  This means that
whenever the tunnel coupling in the Hamiltonian \eqref{eq:2qubits} is switched
off, \textit{i.e.} for $\Delta_1 = \Delta_2 = 0$, the purity decay rate
vanishes.  Thus, we can conclude that the significant purity loss for the
{\sc cnot} operation studied below [cf.\ Eq.~\eqref{cnotgateseq}], stems from the
Hadamard operation.  This is remarkably different from cases with other bath spectra:
For an ohmic bath, for which $J_i(\omega) \propto \omega$, expresion \eqref{Pdot0}
converges in the limit $\omega\to 0$ to a finite value.  By contrast, for a
sub-ohmic bath, this limit does not exist and, consequently, the purity decay
cannot be estimated by its decay rate.
During the stage of the Hadamard operation, $\Delta_2 = \Delta$ while
$\Delta_1=0$.  Then, the interaction picture versions of the qubit-bath
coupling operators read
\begin{eqnarray}
\tilde\sigma_{z,1}(-\tau) &=& \hat \sigma_{z,1} ,\\
\tilde\sigma_{z,2}(-\tau)
&=& \hat \sigma_{z,2}\cos(\Delta\tau/\hbar) - \hat \sigma_{y,2}\sin(\Delta\tau/\hbar) .
\end{eqnarray}
In the case where both qubits couple to individual environments,
the expression for the change of the purity can be evaluated for each
qubit separately.  For qubit 2, we still have a coupling to a good quantum
number, while for qubit 1, the appearence of $\cos(\Delta\tau/\hbar)$ results
in a Fourier integral evaluated at the frequency $\Delta/\hbar$.  Thus, we
finally obtain
\begin{equation}
\dot{\mathcal{P}} =
-\frac{4kT}{5} \lim_{\omega\to 0} \frac{J_1(\omega)}{\hbar\omega}
-\frac{1}{5} J_2(\Delta/\hbar)\coth\frac{\Delta}{2kT}
. \label{eq:exp2}
\end{equation}
For the super-Ohmic bath under consideration [see eqs. (\ref{eq:lowfreq1}) and
\eqref{eq:lowfreq2}], the first term in Eqn. (\ref{eq:exp2}) vanishes.

In the case of one common heat bath, the estimate of the purity decay is
calculated in the same way.  The only difference is that
we have to consider, in addition, cross terms of the
type $\hat \sigma_{1,z} \otimes \hat \sigma_{2,z}$, \textit{i.e.} terms that contain
operators of different qubits.  The contribution of these terms, however,
vanishes when performing  the trace over the bath variables in Eq.~\eqref{eq:exp1}.  Thus, we can
conclude that within this analytical estimate, the purity decay rate is
identical for both the individual bath model and the common bath model.

The so-called quantum degree
\begin {equation}
\mathcal{Q} = \max_{\rho_{\rm out},\ket{\Psi_{\rm me}}}\bra{\Psi_{\rm me}}\rho_{\rm out}\ket{\Psi_{\rm me}}
\end {equation}
is the overlap of the state obtained after the simulated gate operation 
and the maximally entangled Bell states. Finally the entanglement capability $\mathcal{C}$ is 
defined as the smallest eigenvalue of the density matrix resulting from transposing the partial 
density matrix of one qubit. As shown in Ref.~[\onlinecite{peres}], the non-negativity of
this smallest eigenvalue is a necessary condition for the separability of the density matrix
into two unentangled systems. The entanglement capability approaches $-0.5$ for the ideal {\sc cnot} gate.

It has been shown that the controlled-NOT ({\sc cnot}) gate together with single-qubit 
operations is sufficient for universal quantum computation. Here, we investigate the 
decoherence during a {\sc cnot} gate which generates maximally entangled Bell states 
from unentangled input states. In Figures \ref{fig:phonons} and 
\ref{fig:phonons_sc} the simulated gate evolution in the presence of phonon baths is shown. 
Using the system Hamiltonian, the {\sc cnot} gate can be 
implemented through the following sequence
of elementary quantum gates \cite{thorwart,pra_markus}
\begin{eqnarray} \label{cnotgateseq}
U_{\rm CNOT} & = &
U_{\rm H}^{(2)}
\exp \left(-i \frac{ \pi }{4} \hat \sigma_{z,1} \right)
\exp \left(-i \frac{ \pi }{4} \hat \sigma_{z,2} \right) {} \times \nonumber \\
& &{} \times
\exp \left(-i \frac{ \pi }{4} \hat \sigma_{z,1} \hat \sigma_{z,2}\right)
\exp \left(-i \frac{ \pi }{2} \hat \sigma_{z,1}\right)
U_{\rm H}^{(2)}
\textrm{, }\quad \nonumber \\
\end{eqnarray}
where $U_{\rm H}^{(2)}$ denotes the Hadamard gate operation performed on the second qubit.
This gate sequence just involves one two-qubit operation at step three.
The parameters for the numerical calculations are given below Figs. \ref{fig:phonons} and \ref{fig:phonons_sc}.

In Fig.~\ref{fig:phonons}, the gate quality factors for the case of a single or two distinct phononic baths are
shown. It is observed that for the case of a single phonon bath they achieve better values.
This offset is due to the larger number of non-vanishing matrix elements in the coupling of the noise to the spin
components for the two bath case. Here, due to several non-commuting terms in the coupling to the bath and the
different Hamiltonians needed to perform the individual steps of the quantum gate, the gate quality factors
saturate when the temperature $T$ is decreased. This happens at around $T=T_s=12$\,mK corresponding to $E_s=1/4$~GHz as the
characteristic energy scale.

Figure~\ref{fig:phonons_sc}, depicts the same behaviour of the gate quality factors as in Figure~\ref{fig:phonons} with
the only difference that the tunnel coupling $\Delta_2$ is smaller by a factor of $4$ during the Hadamard operation. The qualitative
behavior is very similar to that in Figure~\ref{fig:phonons}, but the deviation from the ideal values for the gate quality
factors is much smaller and already fulfills the criterion of an allowed deviation of $10^{-4}$.
The reduction of the tunnel amplitudes by a factor $4$ corresponds to a very small change of the distance $d$ in
the two qubits (namely, from $100.0$ nm to $100.3$ nm) owing to the Gaussian shape of the electron wavefunctions,
provided their distance is sufficiently large \cite{BrandesHabil}.

We have already mentioned that the phonon contribution to decoherence still allows for the fidelity values
below the threshold $1-F<10^{-4}$ from Ref.\ [\onlinecite{DiVincenzo}].
For a reliable quantum computer, however, such intrinsic decoherence mechanisms
should beat the threshold at least by an order of magnitude. This can be achieved as follows: As we have 
seen, the Hadamard gate is the step limiting the performance as during the Hadamard the system 
is vulnerable against spontaneous emission at a rate $\gamma\propto E^3$, where $E$ is the typical energy splitting of
the single qubit. The duration of the Hadamard, on 
the other hand, scales as $\tau\propto1/E$. Thus, the error probability and the purity decay reduces to $1-e^{-\gamma\tau}\simeq \gamma\tau\propto E^2$.
Thus, by making the Hadamard {\em slower}, {\it i.e.}, by
working with small tunnel couplings between the dots, the gate performance can be increased. This 
works until Ohmic noise sources, electromagnetic noise on the gates and controls, takes over. 
This is demonstrated nicely in Fig.~\ref{fig:phonons_sc}, where the {\sc cnot} gate for a modified Hadamard
opertation (on the second qubit) with $\Delta_2=\varepsilon_2=(1/4)E_s$ is depicted. It is clearly observed that
by decreasing the tunnel matrix element and by increasing the evolution time the decoherence
is reduced and the threshold for the gate quality factors to allow universal quantum computation \cite{Aharonov}
can be achieved.

The gate quality of a CNOT under decoherence has been studied in Refs.\ [\onlinecite{thorwart,pra_markus}] for standard
collective and/or single-qubit noise in Ohmic environments. The single-qubit case for charge qubits in GaAs has
been studied in Ref.\ [\onlinecite{thorwart_mucciolo}] with emphasis on non-Markovian effects. Even in view of this, and
in view of the emphasis of the strong tunneling regime, that work arrives at the related conclusion that intrinsic
phonon decoherence in this system can be limited. Please note, that the approximations in the microscopic model
give an upper bound of validity for the validity of effective Hamiltonians as studied
in Ref. [\onlinecite{thorwart_mucciolo}] as descibed in Refs.\ [\onlinecite{Leonid1,Vorojtsov,Brandes1,BrandesHabil}].
The work presented here is not affected by this restriction due to the emphasis of the case of small tunnel coupling.

\section{Conclusions}

We have analyzed the influence of a phononic environment on four coupled quantum dots which represent
two charge qubits.  The effective error model resulting from the microscopic Hamiltonian does not
belong to the familiar classes of local or collective decoherence. It contains a dipolar and quadrupolar contribution with super-ohmic spectra at low frequencies, $\omega^3$ and $\omega^5$ respectively. The resulting decoherence is an intrinsic limitation of any gate performance.
In particular, we have investigated within a Bloch-Redfield theory the relevant rates and
the quality of a {\sc cnot} gate operation.  The two employed models of coupling the qubits to individual heat
baths versus a common heat bath, respectively, yield quantitative differences for the gate qualifiers.
Still the qualitative behavior is the same for both cases.

Within an analytical estimate for the purity loss, we have found that the decoherence plays its role
mainly during the stage of the Hadamard operation.  The physics behind this is that during all the other
stages, the bath couples to the qubits via a good quantum number.  Consequently, during these
stages, the decoherence rates are dominated by the spectral
density of the bath in the limit of zero frequency which for the present case of a super-ohmic bath vanishes.
The results of our analytical estimate compare favorably with the results from a numerical propagation.

The fact that on the one hand, the bath spectrum is super-ohmic, while on the other hand, the Hadamard operation
is the part that is most sensitive to decoherence, suggests to slow down the Hadamard operation by using
a rather small tunnel coupling.  Then, decoherence is reduced by a factor that is larger than the extension of the
operation time.  This finally results for the complete gate operation in a reduced coherence loss.
Thus, the gate quality is significantly improved for dots with weak tunnel
coupling and can intrinsically meet the threshold for quantum error correction.

\section{acknowledgements}

Our work was supported by DFG through SFB 631. We thank Stefan Ludwig for hinting on the idea
of working with small tunneling and Peter H\"anggi for interesting discussions.

\appendix
\section{Average over all pure states}
\label{app:average}

In this appendix, we derive formulas for the evaluation of expressions of the
type $\mbox{tr}(\rho A)$ and $\mbox{tr}(\rho A\rho B)$ in an ensemble average
over all pure states $\rho=|\psi\rangle\langle\psi|$.  The state
$|\psi\rangle$ is an element of an $N$-dimensional Hilbert space.  Decomposed
into an arbitrary orthonormal basis set
$\{|n\rangle\}_{n=1\ldots N}$, it reads
\begin{equation}
|\psi\rangle = \sum_n c_n |n\rangle ,
\end{equation}
where the only restriction imposed on the coefficients $c_n$ is the normalization
$\langle\psi|\psi\rangle = \sum_n |c_n|^2=1$.  Hence the ensemble of pure
states is fully described by the distribution
\begin{equation}
\label{Pc}
P(c_1,\ldots,c_N) = \gamma_N \delta(1-\sum_n |c_n|^2) .
\end{equation}
We emphasize that $P(c_1,\ldots,c_N)$ is invariant under unitary
transformations of the state $|\psi\rangle$.  The prefactor $\gamma_N$ is
determined by the normalization
\begin{equation}
\int d^2c_1 \ldots d^2c_N\, P(c_1,\ldots,c_N) = 1
\end{equation}
of the distribution, where $\int d^2c$ denotes integration over the real and the imaginary part of $c$.

The computation of the ensemble averages of the coefficients with the
distribution \eqref{Pc} is straightforward and yields
\begin{align}
\overline{c_m c_n^*} = {}& \frac{1}{N}\delta_{mn} \label{c2} \\
\overline{c_m c_n^* c_{m'} c_{n'}^*}
= {}& \frac{1}{N(N+1)}(\delta_{mn}\delta_{m'n'} + \delta_{mn'}\delta_{nm'}) .
\label{c4}
\end{align}
Using these expressions, we consequently find for the ensemble averages of the
expressions $\mbox{tr}(\rho A)$ and $\mbox{tr}(\rho A\rho B)$ the results
\begin{align}
\overline{\mbox{tr}(\rho A)}
={} & \overline{\langle\psi|A|\psi\rangle}
=   \frac{\mbox{tr} A}{N} ,
\\
\overline{\mbox{tr}(\rho A\rho B)}
={} & \overline{\langle\psi|A|\psi\rangle \langle\psi|B|\psi\rangle}
=   \frac{\mbox{tr} (A) \mbox{tr} (B) + \mbox{tr} (AB)}{N(N+1)}
\end{align}
which have been used for deriving the purity decay \eqref{rho2dot} from
Eq.~\eqref{eq:exp1}.

While this averaging procedure is very convenient for analytical
calculations, the numerical propagation can be performed with only a
finite set of initial states.  In the present case, the averages are
computed with the set of 36 states given after Eq.~\eqref{fidelity}.
In the present case, we have justified numerically that both averaging
procedures yield the same results.  Thus, it is interesting whether
this correspondence is exact.

For the case of one qubit, $N=2$, the discrete set of states is given
by the states $|\psi\rangle = c_1|1\rangle + c_2|2\rangle$
where $(c_1,c_2)$ is chosen from the set of 6 vectors
\begin{equation}
\label{c.discrete}
\left(\begin{array}{c} 1\\ 0 \end{array}\right),
\left(\begin{array}{c} 0\\ 1 \end{array}\right),
\frac{1}{\sqrt{2}}\left(\begin{array}{c} 1\\ e^{i\phi} \end{array}\right),
\end{equation}
where $\phi = 0, \pi/2, \pi, 3\pi/2$.  Computing the
averages for the states \eqref{c.discrete} is now staightforward and
shows that this discrete sample also fulfills the relations \eqref{c2}
and \eqref{c4}.  Thus, we can conclude that for the computation of
averages, both the discrete and the continuous sample.

For more than one qubit, however, arises a difference: While the
sample of all pure states also contains entangled states, these are by
construction excluded from set of direct products of the 6 one-qubit
states \eqref{c.discrete}.  Still our numerical results indicate that
the different samples practically result in the same averages.

%---------------------------------------------------------------

%---------------------------------------------------------------

\end{document}